\documentstyle[12pt,aaspp4]{article}
\def\farcs{\hbox{$\> .\!\!^{\prime\prime}$}}

\begin{document}

\def\msa{{MS~1221.8$+$2452}}
\def\msb{{MS~1407.9$+$5954}}
\def\msc{{MS~2143.4$+$0704}}
\def\etal{{\it et~al.}}
\def\eg{{\it e.g.}}
\def\deVauc{{De~Vaucouleurs}}

\title{{\it HST} Imaging of the Host Galaxies of Three X-Ray Selected
BL~Lacertae Objects\altaffilmark{1}
\altaffiltext{1}{Based on observations with the
NASA/ESA {\it Hubble Space Telescope}, obtained at the Space Telescope
Science Institute, which is operated by the Association of
Universities for Research in Astronomy, Inc., under NASA contract
NAS5-26555.}}

\author{Buell T. Jannuzi}
\affil{National Optical Astronomy Observatories, P.O. Box 26732,
Tucson, AZ 85726-6732, USA}
\affil{Email: \tt bjannuzi@noao.edu}

\author{Brian Yanny}
\affil{Fermi National Accelerator Lab., Batavia, IL 60510, USA}
\affil{Email: \tt yanny@sdss.fnal.gov}

\author{Chris Impey}
\affil{Steward Observatory, University of Arizona, Tucson, AZ 85721, USA}
\affil{Email: \tt impey@as.arizona.edu}

\begin{abstract}

{\it Hubble Space Telescope} ({\it HST}) WFPC-2 $I$-band (F814W)
images of three X-ray selected BL~Lacertae objects (\msa, \msb, \&
\msc) reveal that each of these BL~Lac objects is well-centered in an
extended nebulosity that is consistent in brightness and morphology
with being light from an elliptical galaxy at the previously reported
redshifts of these BL~Lac objects.  Each of the detected host galaxies
have radial surface brightness profiles that are well fit by a
\deVauc~law with effective radii of between 3 to 12 kpc ($H_0=50$
km s$^{-1}$ Mpc$^{-1}$, $q_0 = 0$).  The absolute magnitudes of the
host galaxies fall in the range $-24.7 < M_I < -23.5$, in the range of
luminosities determined for other BL~Lacertae object host galaxies.

In addition to allowing the measurement of the host galaxy magnitudes
and radial surface brightness profiles, the {\it HST} images allow a
search for substructure in the host galaxies and the presence of close
companion galaxies at spatial resolutions not yet achievable from the
ground. While no evidence was found for any ``bars'' or spiral arms,
``boxy'' isophotes are present in the host galaxy of at least one of
the three objects observed as part of this study (\msc).  The apparent
magnitudes and image properties of the companions of the BL~Lac
objects are catalogued as part of this work.  The three BL~Lacs appear
to occur in diverse environments, from being fairly isolated (\msa) to
possibly being a member of a rich group of galaxies (\msb).

\end{abstract}


To appear in {\it The Astrophysical Journal Nov 20, 1997}

\section{Introduction}

The spectral energy distributions of catalogued BL~Lacertae objects
are dominated from radio to ultraviolet wavelengths by the variable
and polarized emission of a compact synchrotron source ({\it e.g.}
Impey and Neugebauer 1988; Wagner \& Witzel 1995). The synchrotron
emission is believed to be generated in a ``jet'' of relativistic
material viewed nearly along the jet axis (Blandford \& Rees 1978).
The Doppler boosting of the emission to large apparent luminosities
results in a very bright point source at optical and IR wavelengths
and greatly complicates efforts to observe the nearby environment of
BL~Lac objects.

Observations of the surrounding nebulosities (or host galaxies) and
companions of BL~Lacertae objects are of particular interest because
they provide direct tests of proposed unification schemes between
BL~Lac objects and other classes of AGNs or radio galaxies (\eg~with
Farnaroff-Riley~I radio galaxies, Urry \& Padovani 1995) and allow the
investigation of the role of the nearby environment in generating and
maintaining the observed properties of BL~Lacertae objects.  Despite
the difficulties caused by the bright point sources of BL~Lacs,
studies of the host galaxies of BL~Lacs have been successfully
undertaken from the ground for over 20 years (Oke \& Gunn 1974; Thuan,
Oke, \& Gunn 1975; Ulrich \etal~1975; Miller, French, \& Hawley 1978;
and many others including: Ulrich 1989; Abraham \etal\ 1991; Pesce,
Falomo, \& Treves 1994, 1995; Wurtz, Stocke, \& Yee 1996, hereafter
WSY; Wurtz \etal~1997; Falomo 1996, and references therein). While we
will review some of these past results in \S 3.2, the main result of
the past work can be generalized as follows: whenever the host
galaxies of BL~Lacs have been well observed they have proven to be
elliptical galaxies.

Despite the impressive past results, the majority of well observed
host galaxies are at low redshift ($z<0.6$) or of BL~Lacs that do not
have large ratios of observed nuclear brightness to surrounding
nebulosity.  Studies of BL~Lacs using the high spatial resolution
obtainable with the {\it Hubble Space Telescope} ({\it HST}) have now
been made, with advantages over ground based observations for
measuring the properties of the host galaxy close to the nucleus (a
factor of 2 to 10 times closer depending on the quality of the
ground-based imaging) and for detecting close companions (Falomo
\etal\ 1997; Yanny, Jannuzi, \& Impey 1997).

Unfortunately, the number of BL~Lacs that have been well imaged with
{\it HST} is small and well behind the significant number of luminous
quasars that have been observed (\eg~Disney \etal\ 1995; McLeod \&
Rieke 1995; Bahcall \etal\ 1997, and references therein; Hooper,
Impey, \& Foltz 1997).  Efforts to obtain a significant sample of well
{\it HST} imaged BL~Lacs are in progress by two groups of researchers,
a team led by C. M. Urry and our team. We coordinated target selection
and filter and detector choices in an attempt to provide, in the end,
as large and homogeneous a data set as possible.  The WFPC-2 PC was
chosen as the instrument to use because of its large dynamic range and
pixel scale.  The F814W filter is used in these BL~Lac host
observations to allow comparison with ground based $I_{\rm KC}$-band
fluxes and to maximize the contrast of a host galaxy (expected to be
red in color) over the non-thermal point source.

First results from the Urry team on the imaging of three radio
selected BL~Lacs are presented by Falomo \etal~(1997).  Two of these
objects were found to be in luminous elliptical hosts, while no
underlying host was detected for the third.  We have previously
reported results of our imaging of the radio selected BL~Lac object
OJ~287 (Yanny \etal\ 1997), which might have an off-centered host
galaxy or show evidence of a recent interaction with a close companion
galaxy.  In this paper we present ({\it HST}) WFPC-2 images of three
X-ray selected BL Lacertae objects (XSBLs), \msa, \msb, and \msc, and
discuss what can be learned about their host galaxies and surrounding
environment.  To ease comparison to the analysis of {\it HST} imaging
of BL~Lacs published by Falomo \etal~(1997), we adopt the same
cosmological parameters ($H_0=50$ km s$^{-1}$ Mpc$^{-1}$, $q_0 = 0$)
when deriving absolute luminosities and other physical properties of
the host galaxies.

\section{Observations}

\subsection{Sample Definition}

We selected the three objects discussed in this paper for observation
with {\it HST} because they are all X-ray selected BL~Lac objects in
the complete flux limited {\it Einstein} Observatory Extended Medium
Sensitivity Survey ({\it EMSS}) sample of BL~Lac objects (Morris
\etal~1991). While not evident from the study of
individual objects, some differences in the typical properties of
BL~Lacs selected at different wavelength bands (\eg\ X-ray and radio)
have been noted, particularly in the degree of variability in the
total and polarized optical flux and the polarization position angle
(see Stocke \etal\ 1985, WSY, Jannuzi 1990, and Jannuzi, Smith, \&
Elston 1993, 1994 for examples of further discussion).  Ultimately, if
a larger database of observations can be compiled, we hope to be able
to compare the more isotropic properties of various BL~Lac subsamples
(host galaxy properties, group and cluster environments, etc.).  Some
such comparisons have been made by WSY from their CFHT imaging survey
of 50 objects.  Finding charts and coordinates for the three fields
whose images are presented in this paper may be found in Smith,
Jannuzi \& Elston (1991).

\subsection{WFPC-2 Imaging}

Our observations consisted of multiple long and short exposures with
the WFPC-2 camera of the {\it HST} in the F814W filter.  For
references on the WFPC-2 instrument, see \eg\ Holtzman \etal\ (1995).
For \msa, three 500s and two 200s exposures were obtained on 1996
January 13 UT. For \msb, four 900s and three 160s exposures were taken
on 1995 October 22 UT, and for \msc, two 1100s and five 350s exposures
were recorded on 1995 September 8 UT.  The short exposures were
designed to ensure at least some images of each BL~Lac were obtained
without saturated cores. In practice, none of the three variable
BL~Lac nuclei were bright enough at the time it was imaged to saturate
the detector during even the longest exposures we obtained. Therefore
all exposures of an object were simply averaged together, weighted by
exposure time and with cosmic ray rejection (Yanny \etal~1994).

For each of the three BL~Lac images a background sky level was
subtracted from each image and elliptical isophotes were fit to the
combined image of each object using the ellipse fitting routines of
IRAF/STSDAS.  This allowed us to determine that the unresolved point
source components of each BL~Lac are extremely well-centered in the
surrounding nebulosity.  Comparison of image centroids of the residual
light in the point source function subtracted images shows that the
limits on offsets between the total light (point source plus
nebulosity) and the underlying nebulosities alone are $\delta r <
0\farcs 01$ for \msa, $\delta r < 0\farcs 03$ for \msb, and $\delta r
< 0\farcs 01$ for \msc. These results are consistent with the BL~Lacs
being perfectly centered in each of these cases.

We constructed azimuthally averaged semi-major axis radial profiles
for each object and these are shown in Figure~1.  To aid comparison of
the profiles in Figure~1, we have normalized the central intensity of
the unsaturated point sources at a radius of zero.  There is abundant
excess extended light for each of the BL~Lac objects when compared to
the reference PSF profile seen in Figure~1.  The radii to which the
radial profiles are plotted in Figure~1 were determined by the point
where the signal in the averaged surface brightness becomes comparable
with instrumental read noise in the PC detector.

In Figures 2--4 we show for each of the BL~Lacs the combined WFPC-2 PC
image.  In each figure the target object is in the approximate center.
Objects with measured fluxes are labeled, and the aperture magnitudes
and extents (resolved or point source) are listed in Tables 1--3.  A
large aperture of 150 pixels (6\farcs 9) is used for the host galaxy
and host+PSF measurements quoted in the tables. Apertures with smaller
radii of 6--20 pixels were used for measuring fainter objects in the
field.  The fields were well flattened, and although a small change in
the DC sky level affects the integrated surface brightness and derived
scale radius significantly, the aperture magnitudes are stable to much
better than 0.1 mags over a large range of aperture radii.

Instrumental F814W magnitudes were converted to Kron-Cousins $I$
pass-band using the formulas of Whitmore (1996).  No reddening
correction was applied.  Uncertainties for the separate measurements
of the host galaxy and the point source brightnesses are approximately
0.1 mag due to systematics in the PSF subtraction.  The uncertainty in
the measurement of the combined host+PSF brightness of each object is
much smaller.  The 5$\sigma$ limit on detection for faint point source
objects is approximately $I=24.5$~ in each of the three fields. As in
Figure~1, it is clear in these images that the unresolved point source
components of each BL Lac are surrounded by a resolved nebulosity.

An appropriately scaled model of the WFPC-2 instrumental PSF was
subtracted from each BL~Lac$+$underlying host (Yanny \etal~
1997). Determining the appropriate scaling for the PSF is difficult
for these well-centered objects, and there remains some uncertainty in
the subtraction.  Our normalizations were obtained by examining both
the one dimensional radial profiles and the two dimensional residuals
for a variety of PSF scalings.  Figures~5, 6 and 7 show averaged
semi-major axis plots of the profile of the PSF+host galaxy, host
galaxy alone, and scaled PSF alone for the three objects we studied
for this paper.  Note that in each case, a \deVauc~profile is a much
better fit at small radii than an exponential disk (which would be
expected if the host objects were spiral galaxies), and the extent to
which a \deVauc~profile is a better fit highlights the effectiveness
of the resolving power of {\it HST} over ground-based observations.
We use this evidence in \S 3 as part of the support for identifying
all three host galaxies as elliptical galaxies.

As a check on the accuracy of the model PSF (Krist 1996) which we used
for subtracting the contribution of the unresolved point source to the
image of each BL~Lac, several stellar objects in the field were also
fit and subtracted.  We note that their radial profiles agree well
with that of the model PSF out to a radius where their signal-to-noise
ratio becomes low.  To demonstrate the extent to which the host galaxy
(underlying nebulosity) is present in all three cases, we present in
Figure~8 an image of a field star in the \msb~field, from which a
model PSF has been subtracted.  The BL~Lac in Figure~8 has also had a
scaled PSF subtracted from its image. The residual only contains the
underlying nebulosity, and should be compared to the image in
Figure~3.  The inset in Figure~8 shows \msa~after twice the correctly
scaled PSF has been subtracted in order to show that a mis-scaled PSF
subtraction when extended light is present does not remove light far
from the core. The residual ring is clear evidence for an underlying
host.  No stellar object such as star D in Fig 8, shows a similar
residual.

In Figure~9 we show contour maps, each $4\farcs 6\times 4\farcs 6$,
centered on the three BL~Lacs with the same orientations as shown in
Figures 2--4.  The upper panels show the BL~Lac$+$host galaxy before
subtraction of the scaled PSF, and the lower panel shows the residual
underlying host after a scaled PSF has been subtracted.  The contours
are geometrically spaced by factors of two in flux density, except for
the outer 3 contours which are linearly spaced in flux density.  We
note that the contours for \msc\ appear boxy, further supporting its
interpretation as a large elliptical host.

\subsection{Notes on Individual Objects}

In this section we present additional details about our measurements
of the host galaxies of \msa, \msb, and \msc~and compare our measured
host galaxy properties to some of the existing observations of each
object.  We also make additional notes specific to the individual
object fields. Observed quantities for field objects are listed in
Tables 1--3.  Physical properties of the host galaxies were calculated
assuming $H_0=50$ km s$^{-1}$ Mpc$^{-1}$ and $q_0 = 0$, and are
summarized in Table 4.

\msa~($z=0.218$) has a circular (ellipticity $1-b/a=0.03\pm 0.03$)
host galaxy that is well fit by a \deVauc~profile with a scale length
of 0\farcs68 (3.2~kpc) and $I_{KC}=17.41$.  We measure an $I_{KC}$
surface brightness (SB) at one scale length of $\mu_I = 20.0$ mag
arcsec$^{-2}$.  For the unresolved point source component of the
BL~Lac we measured $I_{KC}=18.06$.  This object was successfully
resolved by WSY during their CFHT imaging survey of BL~Lacs.  They
observed \msa~to contain a point source with Gunn r $>$ 21.2.  They
determined the host galaxy had a brightness in the Gunn~r pass-band of
r$=$18.65 with a surface brightness at one scale length of Gunn~r
$\mu$= 20.82 mag arcsec$^{-2}$. WSY measured a scale length of 0\farcs
6 (2.6 kpc).

\msb~($z=0.495$)  was previously reported by WSY to have an
elliptical host galaxy containing the point source BL~Lac.  We measure
a host with an ellipticity of $1-b/a=0.16\pm0.03$ at P.A. 17$^\circ$, a
\deVauc~scale length of 1\farcs52 (12.2 kpc), and $I_{KC}=18.53$. 
The SB($I$) at one $\rm r_e$ is $\mu_I = 22.3 $ mag arcsec $^{-2}$.
For the BL~Lac itself, we measure $I_{KC}=18.93$.  WSY determined a
host galaxy brightness in the Gunn r pass-band of $r=19.37$, with a
surface brightness at 1 scale length of Gunn r $\mu $=23.4 mag
arcsec$^{-2}$ and a scale length of 1\farcs 4 (10 kpc).  At the time
WSY observed it, the central point source had Gunn $r > 20.1$.

\msc~($z=0.237$) has a host galaxy with an ellipticity 
$1-b/a=0.20\pm0.02$ at P.A. 64$^\circ$, a \deVauc~scale length of
1\farcs79 (9.0 kpc) and $I_{KC} = 17.13$.  The SB($I$) at 1 r$_e$ is
$\mu_I = 21.3 $ mag arcsec$^{-2}$. For the point source, we measure
$I_{KC}=18.58$.  Object L, near the BL~Lac is very elongated, and is a
candidate for a lensed arc. This BL~Lac also has close companions D,
F, and K within a few arcseconds.  WSY quote Gunn $r = 17.89$ for the
large host object, with a SB of $\mu = 22.95$ at one scale radius of
2\farcs 4 (=12 kpc).  The point source had a brightness of Gunn
$r=22.2$ (WSY).

Given the different pass-bands and the variability of the BL~Lacs, our
measurements for the brightnesses and scale lengths of the host
objects agree well with ground based measurements.  Our uncertainties
for the brightnesses of the host galaxies are estimated to be 0.1 mag
(10\%), dominated by systematics in the subtraction of the point
source. To derive quantitative surface brightnesses and effective
radii ($r_e$) measurements, elliptical isophotes were fit to the image
of each host galaxy and our tabulated results correspond to the values
along the semi-major axis. Any deviation from a \deVauc~profile in
the underlying host galaxy (such as the boxy isophotes seen in \msc)
will result in a magnitude reconstructed from the surface brightness
and scale radius which systematically differs from the fixed aperture
magnitude measured for that object. Estimated uncertainties on these
quantities are 0.2 mag on the surface brightnesses and 10\% on the
scale radii.  

The WFPC-2 on {\it HST} provides some of the most accurate
measurements to date of the nebulosity surrounding BL~Lac sources
because of the high dynamic range and spatial resolution provided in
PC images.  These properties allow the separation of PSFs from
extended light at radii as small as $\sim 0\farcs2$.

\section{Discussion}

\subsection{The Host Galaxies are Luminous Elliptical Galaxies}

Our {\it HST} WFPC-2 observations of three X-ray selected BL~Lacs
confirm previous observations of these objects (\eg\ WSY) that
indicated that these BL~Lacs are located in luminous elliptical host
galaxies.  The evidence supporting this conclusion includes the
following:

\begin{itemize}

\item [1.]{The unresolved point source component of each BL~Lac is
extremely well-centered in the surrounding nebulosity or host galaxy
light (see section 2.2), consistent with the point source being
physically associated or ``in'' the nebulosity rather than a
background source being lensed by a foreground galaxy. In this latter
scenario some decentering would be expected (see WSY for discussion and
references therein).}

\item [2.]{The available morphological evidence for each of the host
galaxies strongly supports the classification of these objects as
elliptical galaxies. Both the one dimensional (\deVauc~profile
fitting) and two dimensional appearance of the host galaxies (see
Figures 2--4 and Figure 9) are more consistent with these objects
being elliptical galaxies than spirals.  The least squares fits to the
residual light after subtraction of a scaled point source show a much
better fit by a \deVauc~profile than by an exponential disk profile
(see Figures 5--7).  There is evidence for boxy isophotes in one of
the host galaxies, a characteristic shared by many elliptical
galaxies, while there are no signs in any of the host galaxies of
``bars'', spiral arms, or other substructures that might be expected
in a spiral host galaxy.  The observed ellipticities and scale radii
are also consistent with the properties of elliptical galaxies.}

\item [3.]{The detected host galaxies are luminous.
In Table~4 we list the observed and derived absolute magnitudes for
the three X-ray selected BL~Lac objects in this paper as well as the
two radio selected BL~Lac objects imaged with {\it HST} by Falomo
\etal\ (1997) and the radio and X-ray selected BL~Lac OJ 287 imaged by
Yanny \etal\ (1997).  A K-correction has been applied to each absolute
magnitude based on the catalogued redshift of the BL Lac, our assumed
cosmology, and the typical rest frame colors of ellipticals assumed by
Falomo \etal~(1997) when they made similar calculations to derive the
absolute magnitudes of their detected host galaxies ($V-I=1.3$,
$B-V=0.96$, and $V-R=0.61$).  No correction for the effects of
spectral evolution have been made (as was done by WSY).  If we were to
include the effects of evolution in order to derive what the absolute
magnitudes of these host galaxies would be if they were evolved to the
current epoch, then this might result in a change of from 0.2 mag (for
objects at $z\approx0.2$) to 0.6 mag (for objects at $z\approx0.5$) in
the absolute magnitudes of the objects (see Poggianti 1997 for
examples of the size of the evolutionary corrections required for a
given observed band and assumed intrinsic spectral energy
distribution).  While assuming normal elliptical colors for these host
galaxies is probably not unreasonable, we have only obtained $F814W$
pass-band images with {\it HST}.  Comparing with the ground based
measurements of WSY, the $r - I_{\rm KC}$ colors of the three objects
are 1.2, 0.8 and 0.8 mags respectively.  The derived host galaxy
absolute magnitudes (\eg~ $-23.4 < M_V < -22.2$) are within the range
of observed luminosities of brightest cluster galaxies (BCG), although
perhaps not as bright as the brightest of BCG (the more luminous BCG
have $M_V\approx-23.6$, Hoessel \& Schneider 1985; Postman \& Lauer
1995; see WSY and Hoessel \etal~1980 for distribution of BCG
luminosities).  If spectral evolution is significant, as seems likely,
then since the comparison sample of BCG are typically at redshifts of
0.15, a significant evolution correction would only have to be applied
to the results for \msb. Before comparing the absolute magnitude of
this object's host galaxy to the distribution of BCG absolute
magnitudes we should make fainter the absolute magnitude of \msb~by
0.4 to 0.6 of a magnitude.  As a result, \msb~would be fainter than
the brightest BCG, but still quite a luminous elliptical.}

\end{itemize}

\subsection{Comparison With Other AGN Host Galaxies}

One of the main goals of studying the properties of the host galaxies
of BL~Lacs is to allow a comparison between the observed properties of
the hosts of BL Lacs and the host galaxies of other AGN and candidate
parent populations (\eg~comparing to the observed properties of the
host galaxies and environments of Farnaroff-Riley I and II radio
galaxies; \eg~ Zirbel 1996ab, 1997).  The largest optical survey of
BL~Lac host galaxy properties, ground or space based, is the survey of
made by WSY. They imaged 50 BL~Lacs with the Canada France Hawaii
Telescope and resolved a host galaxy in over 90\% of the objects.
Based on radial profile fits to their images they classified at least
70\% (possibly as high as 90\%) of the detected host galaxies as
ellipticals, with no more than 12\% showing exponential disk or spiral
type hosts.

Despite our small sample size and the resulting inability to draw new
conclusions for the entire class of BL Lac objects, we can place the
objects we observed ``in context'' by comparing the magnitudes of the
hosts of the objects we observed to other well observed BL Lac or
quasar host galaxies.  For example, when our XSBLs are compared to a
sample of Radio Selected BL Lacs studied by Falomo \etal~(1997) at a
comparable redshift, the XSBL host galaxies are found to be about one
magnitude fainter than those of BL Lacs selected by their radio
emission, while still being quite luminous.

We can also compare our BL Lac host galaxies to those observed in the
larger sample of luminous low redshift quasars that have been observed
with WFPC-2, as the nuclear luminosity generating mechanisms may (or
may not) be similar.  Bahcall \etal~(1997) have surveyed the hosts of
20 QSOs with {\it HST} resolution at redshifts $\sim 0.2$.  The host
galaxies of radio quiet QSOs are ellipticals in more than 60\% of the 
cases and have $<M_V> = -22.2\pm 0.6$ on average, about the same 
magnitude as those seen here for XSBL hosts at z=0.2 (see Table 4).  

\subsection{Companion Objects}

The field of the WFPC-2 PC CCD is $36''\times 36''$. At redshift
$z=0.2$, the radius of the field corresponds to $r < 90$~kpc. All
objects within this radius of the three BL~Lacs that are the focus of
this paper were noted in the figures and in Tables 1, 2, and 3. Since
redshifts of these objects are not known, one cannot tell for certain
which objects are true companions to the BL~Lacs, however, objects
within a projected distance of a few arcseconds of the BL~Lac and
brighter than about $I < 24$ are likely to be physically associated
with the BL~Lac.  We note that based on the Hubble Deep Field galaxy 
counts (Williams \etal~1996), the number of galaxies expected 
with $I_{814} < 23.5$ in a circular patch of radius 17$''$ is 
$\sim 4$.

The field of \msa\ has only three objects in the field other than the
BL~Lac+host, and none are within 10$''$.  This appears to be an
isolated BL~Lac.  

In contrast, the field of \msb~is relatively rich, with 20 objects in
the field, all but two of them are non-stellar. \msb\ may be in a
relatively rich group or cluster (see also Wurtz \etal~1993).

The field of \msc\ is intermediate between the other two in the number
of objects in the field, with 12 non-stellar objects.  It has,
however, several very close neighbors which are likely to be
physically associated with \msc. The objects D and F are especially
likely to be associated with the BL~Lac.  Object L is of interest
because its elongation and alignment is reminiscent of gravitationally
lensed objects seen in the field of massive clusters.  A redshift of
object L and further analysis are needed.  We note that {\it HST} has
provided images with similarly elongated objects that have proven to
be spiral galaxies.

Our three randomly selected XSBLs show differing field
environments. This demonstrates clearly, if not unexpectedly, that a
sample of three objects is not large enough to draw any conclusions
about the role that environment may play in the BL~Lac phenomenon.
The richness of the environments of larger samples of BL Lac objects
has been quantitatively studied using ground-based imaging data ({\it
e.g.} Wurtz \etal~1997, including data on the three objects studied
in this paper), but such studies still lack spectroscopic confirmation
of the association with the BL Lacs of the apparent companion objects.
Spectroscopic follow-up would be valuable not only to confirm group
membership, but also to determine velocity dispersions for the groups.

\section{Summary}

We have presented {\it HST} WFPC-2 $I$-band (F814W) images of three
X-ray selected BL~Lacs, measured the properties of the BL~Lac host
galaxies, and catalogued the companion objects in the fields of these
BL~Lac objects. From these data we have determined the following:

\begin{itemize}

\item [1.]{The host galaxies of the X-ray selected BL Lacs 
\msa, \msb, and \msc~ are luminous elliptical galaxies with absolute
luminosities in the range $-24.7 < M_I < -23.5$). These host galaxies
appear typical of the BL Lac host galaxies that have been previously
observed.}

\item [2.]{The three XSBLs observed in our program have a wide range
in the apparent number of companion objects, but without spectra or
color information to allow a more accurate determination of which
objects are true companions of each BL~Lac we are not able to further
investigate whether or not the larger scale environments of these
three objects are disparate nor are we able to compare their
environments to those of other BL~Lacs studied by others ({\it e.g.}
Wurtz \etal~1997).}

\item [3.]{The absolute magnitudes of the BL Lac hosts near z=0.2 have
absolute magnitudes comparable to those of radio quiet QSO hosts seen in 
Bahcall \etal\ (1997). There is a suggestion that XSBL host galaxies
might be  fainter than RSBL hosts, but this needs to be tested with
high quality imaging of larger and well defined samples. }

\end{itemize}

The small sample of 6 objects that have been well imaged by {\it HST}
(this paper; the two objects with detected host galaxies in Falomo
\etal~1997; and OJ~287, Yanny \etal~1997) is not large enough to draw
strong conclusions about ``all'' BL~Lacs or differences in
sub-groupings.  Fortunately additional objects are being imaged and
the total sample size will certainly grow.  {\it HST} imaging offers
the potential of undertaking studies of higher redshift objects that
can be integrated into the larger data sets of images of low redshift
BL~Lacs.

We acknowledge useful discussions with A. Dey, R. Falomo, T. Lauer,
S. Kent, R. Scarpa, and J. Stocke.  We acknowledge support from NASA
grant GO-5992.02-94A.  B.Y. acknowledges support from the Fermi
National Accelerator Laboratory.  B.T.J.  acknowledges support from
the National Optical Astronomy Observatories, operated by the
Associated Universities for Research in Astronomy, Inc., under
cooperative agreement with the National Science Foundation.

\vfil\eject
\centerline{\bf FIGURE~CAPTIONS}

\figcaption{  
Averaged semi-major axis profiles of three BL Lac objects plus a
bright model field star to show the PSF.  All three of the BL Lacs
clearly show extensive extended light in excess of that expected from
an isolated point source.
\label{fig1}}

\figcaption{
Image obtained with the WFPC-2 PC of the \msa~field.  The BL~Lac and
host is in the center of the image. No subtraction has been performed.
Measurements for labeled objects are listed in Table~1.
\label{fig2}}

\figcaption{
Image obtained with the WFPC-2 PC of the \msb~ field.  The BL~Lac and
host is in the center of the image. No subtraction has been performed.
Measurements for labeled objects are listed in Table~2.  Note the rich
field.
\label{fig3}}

\figcaption{
Image obtained with the WFPC-2 PC of the \msc~ field.  The BL~Lac and
host is in the center of the image. No PSF subtraction has been
performed.  Measurements for labeled objects are listed in Table~3.
Note the close companions D and L. L is very elongated.  The apparent
``ejecta'' from the SE of the object is a diffraction spike produced
by the central point source.
\label{fig4}}

\figcaption{
Elliptical isophote fits to the image of \msa.  The upper solid line
shows the radial profile (elliptical semi-major axis) of the
sky-subtracted data.  The lower solid line is a {\it HST} stellar
point spread function scaled to match the point source component of
the BL Lac image.  The asterisks mark the underlying host galaxy light
once the scaled point source has been centered and subtracted from the
image.  The filled circles represent the best least chi-squared fit
for a \deVauc~profile to the host galaxy light for radii between
0\farcs 15 and 1\farcs 5.  The open circles represent the best fit
exponential disk profile to the host galaxy light.  The small error
ticks at 1\farcs 6 and 2\farcs 1 show read noise induced errors in the
surface brightness measurements.
\label{fig5}}

\figcaption{
Same as Fig. 5, but for \msb.
\label{fig6}}

\figcaption{
Same as Fig. 5, but for \msc.
\label{fig7}}

\figcaption{
In this image are displayed the results of subtracting a scaled PSF
from several objects in our images. In the main image, the Galactic
field star labeled D, has had a PSF scaled and subtracted from it.
Note the complete lack of any diffuse emission beyond the core of the
residual.  This is the expected result for all point source objects.
In contrast, after subtraction of a scaled PSF from \msb~ (shown in
the center of the image), there is considerable residual extended
emission from the elliptical host galaxy. In the inset, a PSF of twice
the fitted value was intentionally over-subtracted from \msa\ to show
the extended emission in contrast to the lack of emission surrounding
the field star D in the main field.
\label{fig8}}

\figcaption{
Contour plots of the images of the central $4\farcs 6 \times 4\farcs
6$ boxes of each of the three BL Lac objects before (upper panels) and
after (lower panels) scaled PSF subtraction.  Each of the point
sources is well-centered on its host galaxy.  The isophotes for
object \msc\ appear boxy, additional evidence that this host galaxy is
an elliptical.  The ring of circles seen in \msa\ after PSF
subtraction are due to slight changes in the {\it HST} diffraction
level optics between the BL~Lac and the model PSF.
\label{fig9}}

% Start of tables

\begin{table}
\caption{Objects in the Field of  \msa \label{table1}}
\begin{tabular}{rrrrr}
\tableline
\multicolumn{1}{c}  {ID} &
\multicolumn{1}{c} {$\Delta$RA($''$)} &
\multicolumn{1}{c} {$\Delta$DEC($''$)} &
\multicolumn{1}{c} {$I_{KC}$ mag} &
\multicolumn{1}{c}  {FWHM($''$)\tablenotemark{a}} \cr
\tableline
\msa &   0.00&    0.00&   16.94 & ext \cr
HOST &   0.00&    0.00&   17.41 & 0.68\tablenotemark{b} \cr
POINT SOURCE &   0.00&    0.00&   18.06 & pt \cr
 A & $-$8.12& $-$12.41&   21.24 & 0.14 \cr
 B &$-$10.52&    9.10&   21.41 & 0.14 \cr
 C &  13.60&  $-$2.54&   22.27 & ext \cr
\end{tabular}
\tablenotetext{a}{The label ``ext'' indicates the object is
significantly extended, while ``pt'' indicates a point source.}

\tablenotetext{b}{For the host galaxy (HOST), FWHM is replaced by the
elliptical 
scale radius r$_e$.} 
\end{table}

\begin{table}
\caption{Objects in the Field of  \msb~ \label{table2}}
\begin{tabular}{rrrrr}
\tableline
\multicolumn{1}{c}  {ID} &
\multicolumn{1}{c} {$\Delta$RA($''$)} &
\multicolumn{1}{c} {$\Delta$DEC($''$)} &
\multicolumn{1}{c} {$I_{KC}$ mag} &
\multicolumn{1}{c}  {FWHM($''$)\tablenotemark{a}} \cr
\tableline
\msb & $-$0.00&  $-$0.01&   17.96 & ext \cr
HOST & $-$0.00&  $-$0.01&   18.53 & 1.52\tablenotemark{b} \cr
POINT SOURCE &   0.29&    0.16&   18.93 & pt \cr
 A & $-$8.30&  $-$9.02&   :17.81 & ext \cr
 B & $-$7.40&  $-$5.74&   20.68 & 0.11 \cr
 C &   0.18& $-$11.15&   20.99 & 0.09 \cr
 D &  10.58& $-$10.98&   18.60 & pt \cr
 E &   5.22&    2.91&   21.18 & 0.12 \cr
 F &$-$12.78&    0.99&   21.69 & 0.17 \cr
 G &  11.21&  $-$7.83&   21.51 & 0.09 \cr
 H &$-$14.19&    6.31&   22.69 & 0.11 \cr
 I &$-$20.91&  $-$3.37&   22.13 & 0.1 \cr
 J &$-$22.39&  $-$0.84&   :21.47 & ext \cr
 K &   8.45&    9.52&   22.78 & 0.27 \cr
 L &   9.91&    6.93&   23.69 & 0.11 \cr
 M &  16.00&    1.78&   23.35 & 0.15 \cr
 N &   2.95& $-$16.39&   23.52 & pt \cr
 O & $-$6.93&  $-$1.94&   24.30 & 0.09 \cr
 P & $-$3.21&  $-$5.32&   24.39 & 0.34 \cr
 Q & $-$5.16&    2.65&   24.91 & ext \cr
 R &$-$10.46&    4.60&   23.85 & 0.17 \cr
 S &   3.02&    5.33&   24.70 & 0.13 \cr
 T &   5.44& $-$13.39&   24.99 & ext \cr
\end{tabular}
\tablenotetext{a}{The label ``ext'' indicates the object is
significantly extended, while ``pt'' indicates a point source.}

\tablenotetext{b}{For the host galaxy (HOST), FWHM is replaced by the
elliptical scale radius r$_e$.} 
\end{table}

\begin{table}
\caption{Objects in the Field of  \msc \label{table3}}
\begin{tabular}{rrrrr}
\tableline
\multicolumn{1}{c}  {ID} &
\multicolumn{1}{c} {$\Delta$RA($''$)} &
\multicolumn{1}{c} {$\Delta$DEC($''$)} &
\multicolumn{1}{c} {$I_{KC}$ mag} &
\multicolumn{1}{c}  {FWHM($''$)\tablenotemark{a}} \cr
\tableline
\msc &   0.00&  $-$0.00&   16.88 & ext \cr
HOST &   0.00&  $-$0.01&   17.13 & 1.79\tablenotemark{b} \cr
POINT SOURCE &   0.00&  $-$0.01&   18.58 & pt \cr
 A & $-$6.79&  $-$9.87&   18.88 & pt \cr
 B &$-$15.87&  $-$3.29&   20.10 & 0.02 \cr
 C &$-$15.92&    9.48&   22.09 & ext \cr
 D & $-$1.31&    0.99&   23.60 & 0.18 \cr
 E & $-$2.38&    6.34&   23.14 & 0.09 \cr
 F & $-$3.97&    1.64&   23.37 & 0.03 \cr
 G &$-$12.95&   13.56&   23.33 & 0.09 \cr
 H &$-$13.67&    9.21&   24.19 & 0.05 \cr
 I & $-$5.28&   14.66&   23.71 & ext \cr
 J &   1.35&    6.17&   24.57 & pt \cr
 K & $-$0.62&    3.28&   25.63 & ext \cr
 L & $-$0.47&  $-$2.56&   24.97 & ext \cr
 M &$-$18.75&   13.62&   24.48 & ext \cr
 N &   4.72& $-$11.56&   23.65 & ext \cr
 O &   2.40&  $-$7.75&   24.87 & ext \cr
\end{tabular}
\tablenotetext{a}{The label ``ext'' indicates the object is
significantly extended, while ``pt'' indicates a point source.}

\tablenotetext{b}{For the host galaxy (HOST), FWHM is replaced by the
elliptical scale radius r$_e$.} 

\end{table}

\begin{table}
\caption{Magnitudes of Selected BL Lacertae Object Host Galaxies \label{table4}}
\begin{tabular}{rrllrlrrr}
\tableline
\multicolumn{1}{c}  {Name} &
\multicolumn{1}{c} {$I$} &
\multicolumn{1}{l} {redshift} &
\multicolumn{1}{l} {k$_{\rm corr}$} &
\multicolumn{1}{l} {m--M} &
\multicolumn{1}{c} {$M_{I}$\tablenotemark{a}} &
\multicolumn{1}{c} {$M_{R}$\tablenotemark{b}} &
\multicolumn{1}{c} {$M_{V}$\tablenotemark{b}} &
\multicolumn{1}{c} {$\rm r_{\rm e}$ (kpc)\tablenotemark{a}} \cr
\tableline
1823+568\tablenotemark{c} & 18.3  & 0.664 & 0.54 &  43.72 & $-$26.0 & $-$25.3 & $-$24.7 & 6.5\cr
2254+074\tablenotemark{c} & 15.9  & 0.19  & 0.11 &  40.49 & $-$24.7 & $-$24.0 & $-$23.4 &15.0\cr
\msa                      & 17.41 & 0.218 & 0.13 &  40.81 & $-$23.5 & $-$22.8 & $-$22.2 & 3.2 \cr
\msb                      & 18.53 & 0.495 & 0.37 &  42.84 & $-$24.7 & $-$24.0 & $-$23.4 &12.2 \cr
\msc                      & 17.13 & 0.237 & 0.14 &  41.01 & $-$24.0 & $-$23.3 & $-$22.7 & 9.0 \cr
OJ~287\tablenotemark{d}   & 18.3  & 0.306 & 0.19 &  41.65 & $-$23.5 & $-$22.8 & $-$22.2 &--\cr \tableline \end{tabular}
\tablenotetext{a}{$H_0=50$ km s$^{-1}$ Mpc$^{-1}$, $q_0 = 0$, $M_I = I
- $(m--M) -- K-corr } 
\tablenotetext{b}{$V-I = 1.3$, $V-R=0.6$, are
assumed, as in Falomo \etal~(1997)} 
\tablenotetext{c}{The radio
selected BL Lac host data from Falomo \etal~(1997).}
\tablenotetext{d}{OJ~287 BL Lac host data from Yanny \etal~(1997).}
\end{table} 
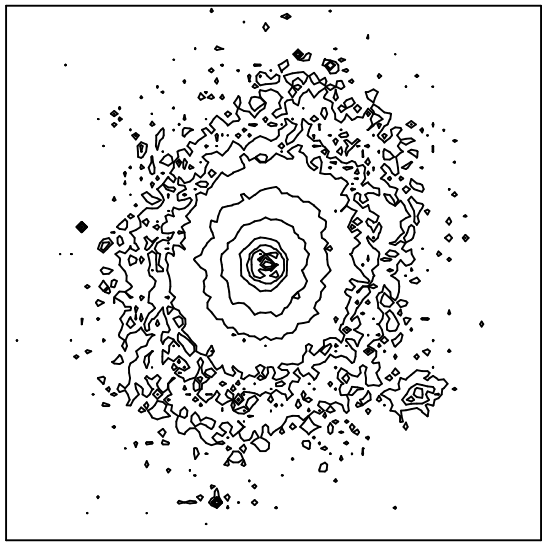
\end{document}